\documentclass{camera}
\usepackage{floatflt,graphicx}
\newcommand\as{\alpha_{\mathrm{S}}}

\begin{document}

%

\renewcommand{\thefootnote}{\fnsymbol{footnote}}

\title{HADRONIC PHYSICS AND QCD: \\  STATUS AND FUTURE \footnote{Summary talk of the Hadronic Physics session at ``IFAE'', the XIV italian meeting of High Energy Physics, Parma April 2002}}

%
\author{M. GRAZZINI$^{a}$ \And R. NANIA$^{b}$}

%
\organization{$^{a}$ Dipartimento di Fisica, Universit\`a di Firenze and INFN, Sezione di Firenze, I-50019 Sesto F., Florence, Italy\\ 
$^b$ INFN Sezione di Bologna, I-40126 Bologna, Italy}

\maketitle

%
\begin{center}
\bf{Abstract}
\end{center}
\begin{quote}
\pretolerance 10000
Studies of hadronic final states are entering a new
phase where very precise experimental measurements require
better theoretical predictions
for a meaningful comparison. 
Recent results and future developments are briefly reviewed
both for the experiments and the theory.
\end{quote}

\renewcommand{\thefootnote}{\arabic{footnote}}
\setcounter{footnote}{0}

\section{Introduction}
Since quite some time the experimental results 
have supported
Quantum Chromo Dynamics (QCD) as the theory of strong interactions.
In recent years
the increasing precision of the experiments
has prompted
a more detailed comparison between
data and theoretical predictions.

The thirteen talks given
in the hadronic physics
session of this conference stimulated an extremely lively discussion
between theorists and experimentalists. 
On the experimental
side particular
attention
is currently devoted 
to the
development of
techniques which could bring to
a cleaner comparison
between data and QCD predictions.
From the theoretical
point of view the main efforts are towards
the extension of the present accuracy
of perturbative calculations and a better understanding
of the hadronization (non-perturbative) regime.

In this short summary only few of the many interesting topics
discussed will be covered. For more details we refer the reader to
the other contributions to these proceedings.

\section{Status of measurements.}

\begin{figure}
\centering
\includegraphics[width=6.0cm]{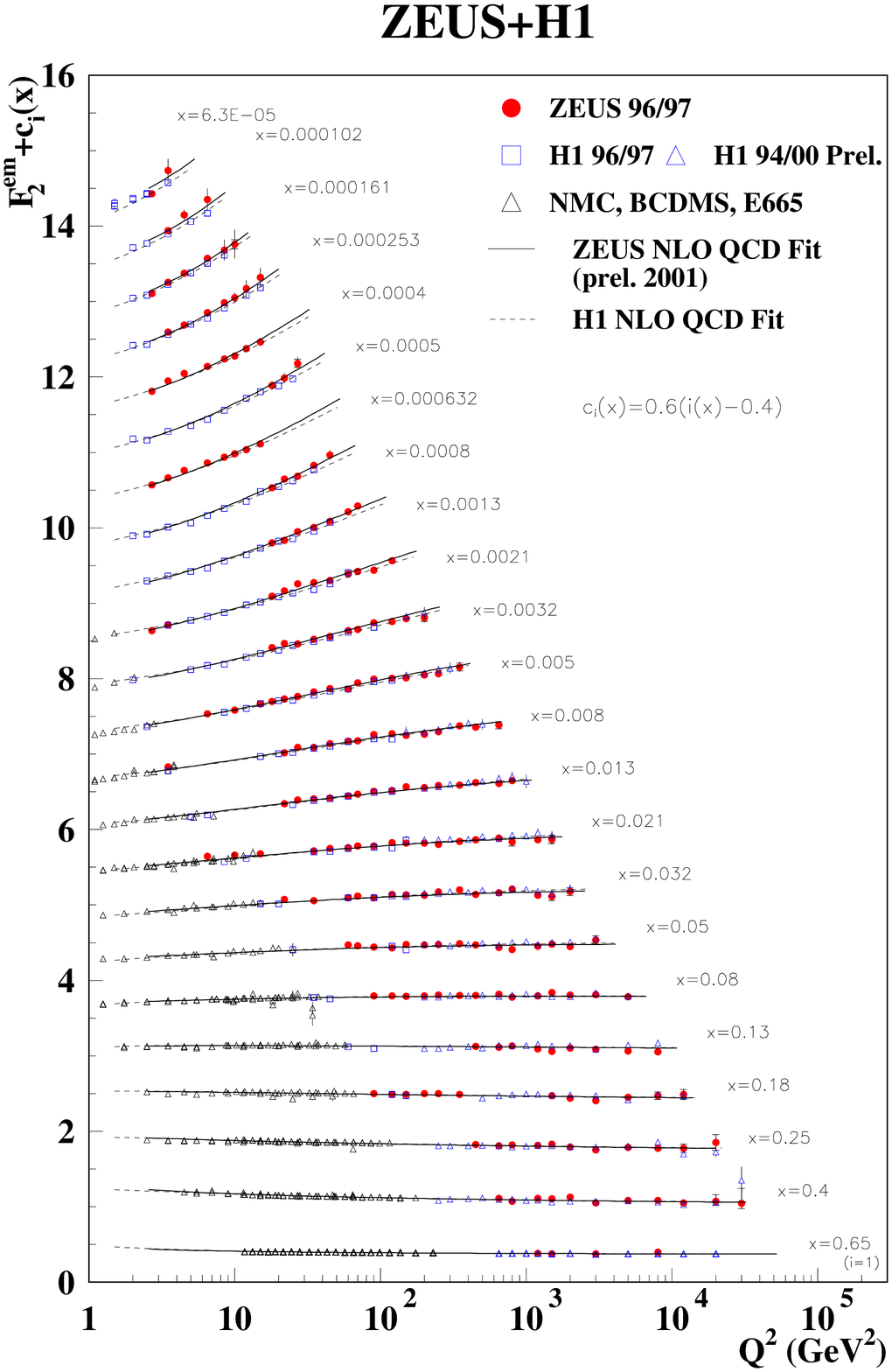}
\includegraphics[width=6.0cm]{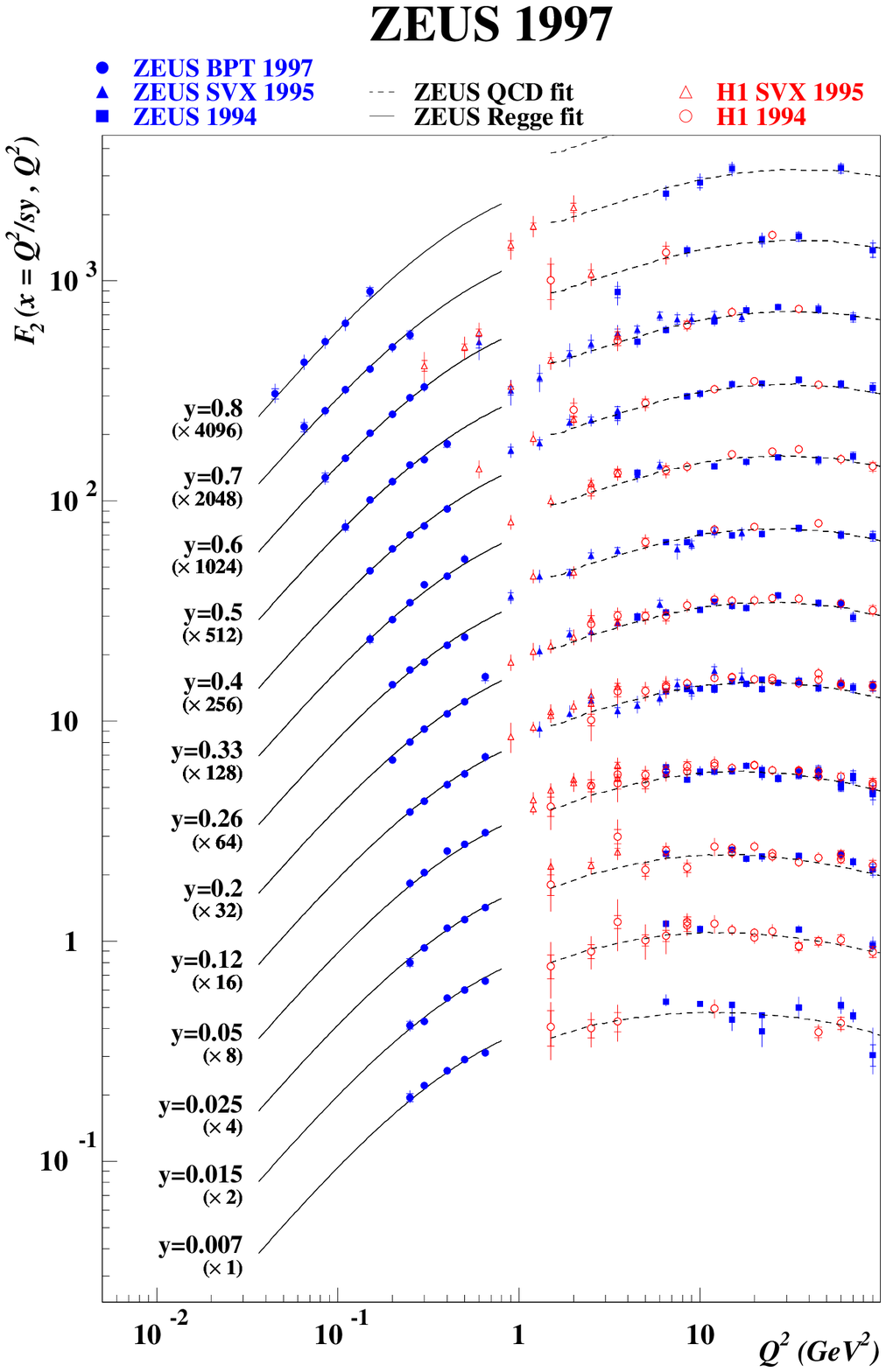}
\caption{Measurements of the proton Structure Function $F_2$
as a function of $Q^2$ for the intermediate/high (left) and very low (right)
$Q^2$ regions.  $x$ is the
Bjorken variable, 
$y$ is the inelasticity in DIS events. } 
\label{fig:f2}
\end{figure}

QCD started with Deep Inelastic Scattering analysis.
Still now the large amount of Structure 
Function measurements ($F_2$)
at HERA provide the best probe to parton 
distribution functions (PDFs) inside the proton and an invaluable test
for QCD  predictions \cite{pellegrino}. A compilation of 
the available data is reported in fig.~\ref{fig:f2}.
Next-to-leading order (NLO) DGLAP fits to these data provide measurements of
the gluon density in the proton if the
value of the
QCD coupling $\as$ 
is given as input, but first attempts
for a combined extraction of the two quantities are 
already available with the following results:
$\as (M_Z)$ = $0.1150\pm 0.0017 (exp) \pm^{0.0009}_{0.0005}(mod) \pm 0.005 (scale)$ 
\cite{h1as} 
and $\as (M_Z)$ = $0.1166\pm 0.0008(uncor)\pm 0.0032(corr) \pm 0.0036 (norm)$
 \cite{zeas}
which demonstrate 
the remarkable precision reached.

In the last years much interest has been raised by precise measurements
of $F_2$ in the very low $Q^2$ region down to 0.1 GeV$^2$.
These data (fig.~\ref{fig:f2} right) \cite{f2low} 
have stimulated 
many theoretical works, the challenge being to extend
QCD predictions
towards this kinematical region,
that cannot be reached
by standard perturbative approaches, and to allow a connection
between Regge phenomenology and QCD calculations \cite{f2lowqcd}.
\begin{floatingfigure}[l]{7.0cm}
\centering
\includegraphics[width=7.0cm]{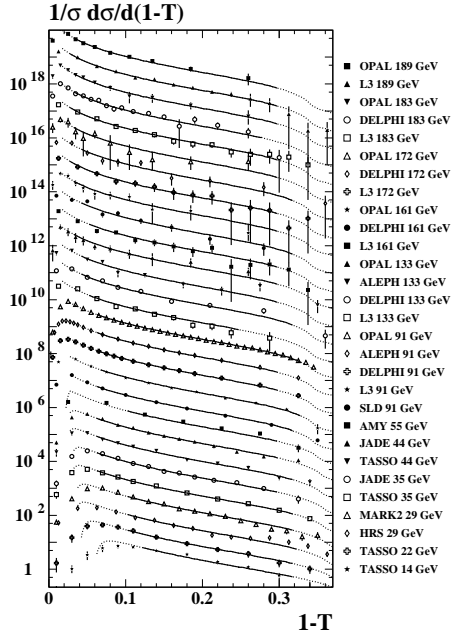}
\caption{The distribution $(1-T)$ as measured from PETRA to LEP.} 
\label{fig:LEP}
\end{floatingfigure}
Although now closed,
LEP is still producing many interesting 
results for our understanding of hadronic physics \cite{fabbri}.
Among these the measurements of $\as$ have reached 
a remarkable precision: thanks
to the various energies scanned
it was possible to study its running
from PETRA energies up to 208 GeV.
An example of the impressive amount of data available is given in
fig.~\ref{fig:LEP}, 
where the distributions for the thrust variable are reported
from different experiments and energies. 
The data are fitted in terms of
analytical calculations which rely on two parameters:
the QCD coupling $\as$ and its non-perturbative extrapolation $\alpha_0$,
which is assumed to be universal \cite{DMW}.
By using five different shape variables this approach
provides the following measurements: 
 $\as(M_Z)$ = $0.1171 \pm^{0.0032}_{0.0020}$ and
 $\alpha_0$ = $0.513 \pm^{0.066}_{ 0.045}$ \cite{LEP}.
The QCD coupling constant is measured precisely, 
and the approximated universality of $\alpha_0$ is confirmed, within the uncertainties.
The accuracy of the measurement is definitely limited by the theory.

Also the ep collider HERA
has now started to produce $\as$ measurements using different 
jet analysis, both in photoproduction and in DIS \cite{lupi}.
A compilation of $\as$ determinations is given in 
fig.~\ref{fig:heraas}, compared
with the world average: the precision reached 
is comparable to other measurements. 
It should be noted that all these studies require high
transverse energy $E_T$ of the jets
or high $Q^2$ of the events in the case of 
the DIS samples. 
\begin{floatingfigure}[l]{4.5cm}
\centering
\includegraphics[width=4.0cm]{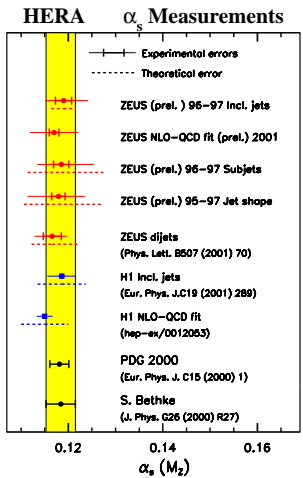}
\caption{$\as$ measurements at HERA from jet analysis.} 
\label{fig:heraas}
\end{floatingfigure}
It would have been interesting also to use the data
to measure the gluon distribution using 
lower $Q^2$ events to reach interesting low  $x$ intervals to compare with
the standard Structure Function analysis. 
In this kinematical region the experiments 
have reached a good precision,
but unfortunately the NLO QCD prediction
has a large theoretical uncertainty,
making the 
extraction of QCD parameters not reliable
\cite{herajetlowq2}. 

Also the CDF collaboration at the Tevatron has recently produced
a first measurement of $\as$ from high-$E_T$ jet data \cite{CDF}.
Several efforts are going on both from the experimental 
and theoretical side to study
how jet physics
at hadron colliders
is affected by
beam-spectator effects, 
initial-final state radiation and multiple parton scattering
and to find possible
alternatives
to the standard jet algorithms \cite{snowmass}.
These studies are fundamental for the present Run II data
and, on longer time scale, for the LHC analysis.

\section{Theoretical challenges and tools}

The previous discussion clearly indicates that the theoretical uncertainty in
the measurement of many QCD observables is becoming dominant.
The present accuracy of fixed order calculations for jet observables
is limited to NLO. It is important to remind that
these calculations are affected by soft and collinear singularities.
Although cancelling in infrared-safe observables, these 
singularities separately affect virtual and real contributions
and are usually handled with general algorithms
that combine the analytic calculation of the singular part
with numerical integration.
The step forward to next-to-next-to-leading order (NNLO) is non trivial,
because, besides involving
the calculation of the relevant two-loop amplitudes,
it requires the full understanding of
the pattern of cancellation
of infrared singularities at ${\cal O}(\as^2)$.
Considerable progress has been recently achieved in 
this direction: important two-loop amplitudes have been 
computed and the singular behaviour of tree and 
loop amplitudes at ${\cal O}(\as^2)$ has been understood \cite{oleari}.
Hopefully this progress will make NNLO calculations feasible
in the next future.
To obtain reliable predictions for hadron colliders,
precise knowledge of PDFs is needed.
In particular, a consistent NNLO calculation 
requires NNLO PDFs. Recently, a PDF set including an approximated NNLO fit
has been produced \cite{mrst2001}.
Another important subject of theoretical investigation is the one
of
trying to quantify the
PDF uncertainties.
Recently, PDF sets with a systematic
study of correlated errors have been released \cite{pdfunc,zeas}:
it is thus possible to estimate
the ensuing uncertainty on the cross section for the relevant
processes at the Tevatron and the LHC.

Remaining on the perturbative side other presentations at this conference reported theoretical progress in soft-gluon resummation \cite{banfi} and in the study of BFKL effects at hadron colliders \cite{vacca}.

Even though it is important to extend the present accuracy of perturbative calculations,
it is clear that it does not make sense
to try pushing indefinitely perturbation theory. 
The theoretical uncertainty affecting a QCD observable
is given not only by the missing higher order contributions
but also by the hadronization corrections
that are often of the same order of magnitude.
In recent years a great effort has been devoted in estimating
hadronization corrections by using perturbative driven approaches,
which lead to power suppressed
effects.
The
best known is certainly the one of Ref.~\cite{DMW}
which is based on the definition of an infrared
finite coupling $\alpha_0$, which is then fitted to the data (see Sect.~2).
The hadronization correction results in a power suppressed
shift of the perturbative distribution.
A more recent and sophisticated approach is based on the
so called ``shape function'',
which is
a non-perturbative quantity
assumed to smear the perturbative distribution \cite{magnea}. 

Besides ``pure'' theoretical progress, great importance has the way in which the current understanding of the theory is implemented in the tools the experimentalists have to their disposal.
In this respect, a great work is being carried out to improve
current Monte Carlo (MC) event generators.
A fixed order perturbative calculation correctly 
incorporates the radiation of hard partons, through the corresponding 
QCD matrix elements, but fails to reproduce soft and collinear 
regions, where multiple QCD emissions become important.
Moreover fixed order perturbative calculations
are not directly suitable to implement hadronization. 
On the other hand a standard MC parton shower correctly 
treats soft and collinear radiation and can incorporate 
hadronization models but does not take into account hard emissions.  
Thus there are efforts to implement hard matrix element 
corrections in the current event generators, 
so as to correctly include the effect of parton emissions
in the full phase space.
In particular multijet matrix elements are being 
implemented in ordinary MC and will provide a more 
reliable tool to study backgrounds for many new 
physics processes at hadron colliders \cite{moretti}.
However it would be of great importance to combine MC with
NLO predictions,
that are the standard in fixed order calculations.
A recent proposal has been described at this conference 
and encouraging results have been presented \cite{frixione}.

\section{The Heavy Flavour Puzzles}
\begin{floatingfigure}[l]{6.0cm}
\centering
\includegraphics[width=6.0cm]{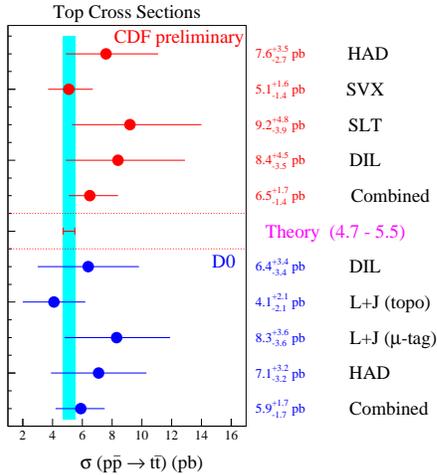}
\caption{Summary of Tevatron  measurements of 
top production and comparison
with theoretical estimates.} 
\label{fig:cdf}
\end{floatingfigure}

The updated $t\bar{t}$ cross section from the CDF collaboration
shows an agreement, within the present statistics,
with the theoretical QCD predictions with a top mass of 174 GeV/c$^2$.
(fig.~\ref{fig:cdf}).

Still unclear instead is the situation with the inelastic $J/\psi$
production \cite{bertolin}. The latest ZEUS results in ep collisions indicates that
NLO color singlet predictions are enough to explain the data and
there is no need for colour octet component as requested by
the CDF data.

The most puzzling situation is certainly the one of beauty production, where there are often large discrepancies between data and theoretical QCD predictions.
This is a long standing problem, found already in the eighties
with the anomalous high heavy flavour
cross section measured 
at the Intersecting Storage Ring \cite{ISR}
and at the $Sp{\bar p}S$ \cite{UA1}.
The CDF collaboration
has recently reported an excess of a factor $2.9$ of
the data over the QCD prediction \cite{sidoti}.
A recent reanalysis, which makes use of all the
theoretical information available, and, in particular, of a more careful treatment
of the fragmentation, brings the 2.9 CDF discrepancy to 1.7 \cite{nason}.

\begin{figure}[h]
\centering
\includegraphics[width=5.5cm]{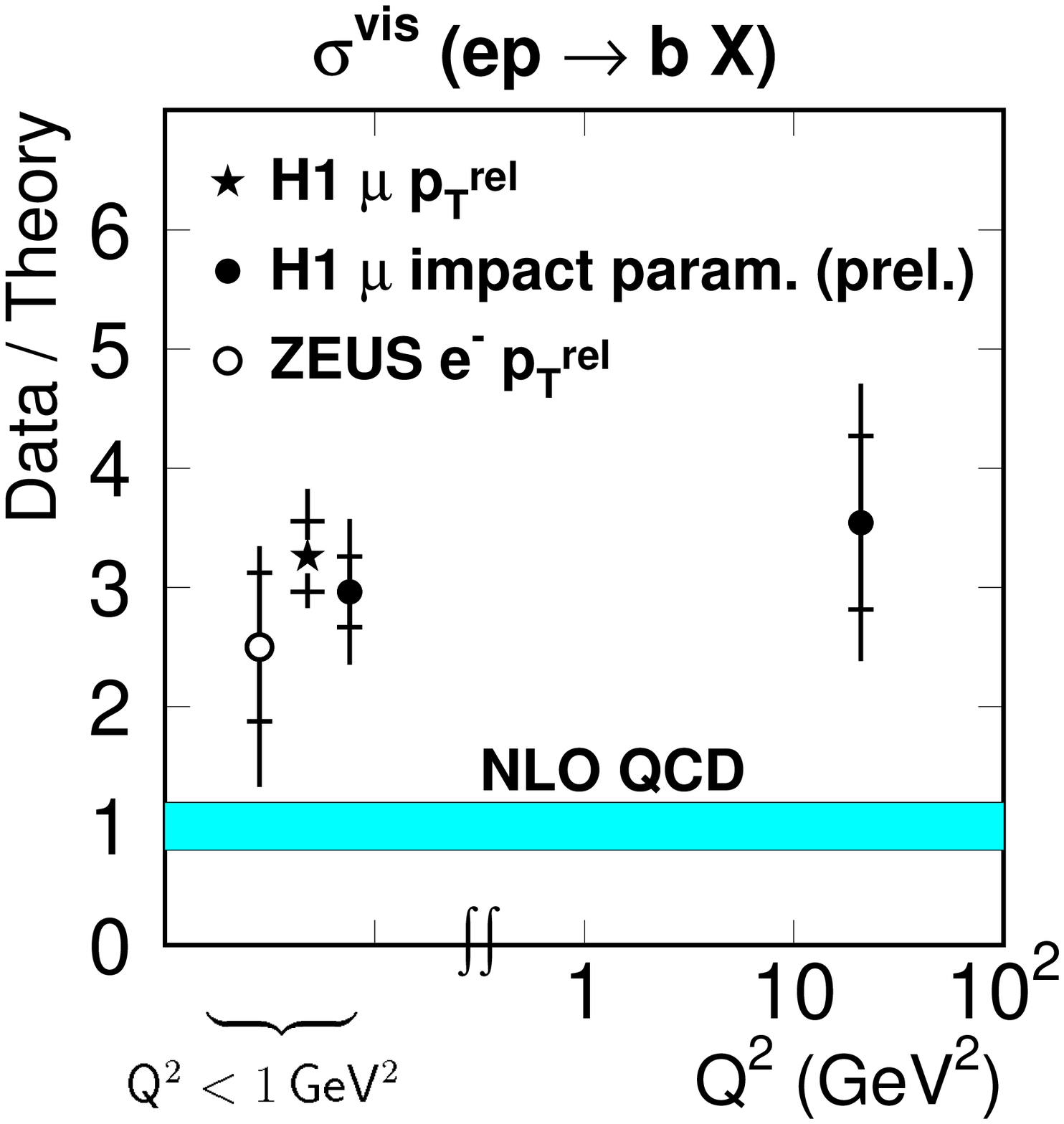}
\includegraphics[width=6.0cm,clip]{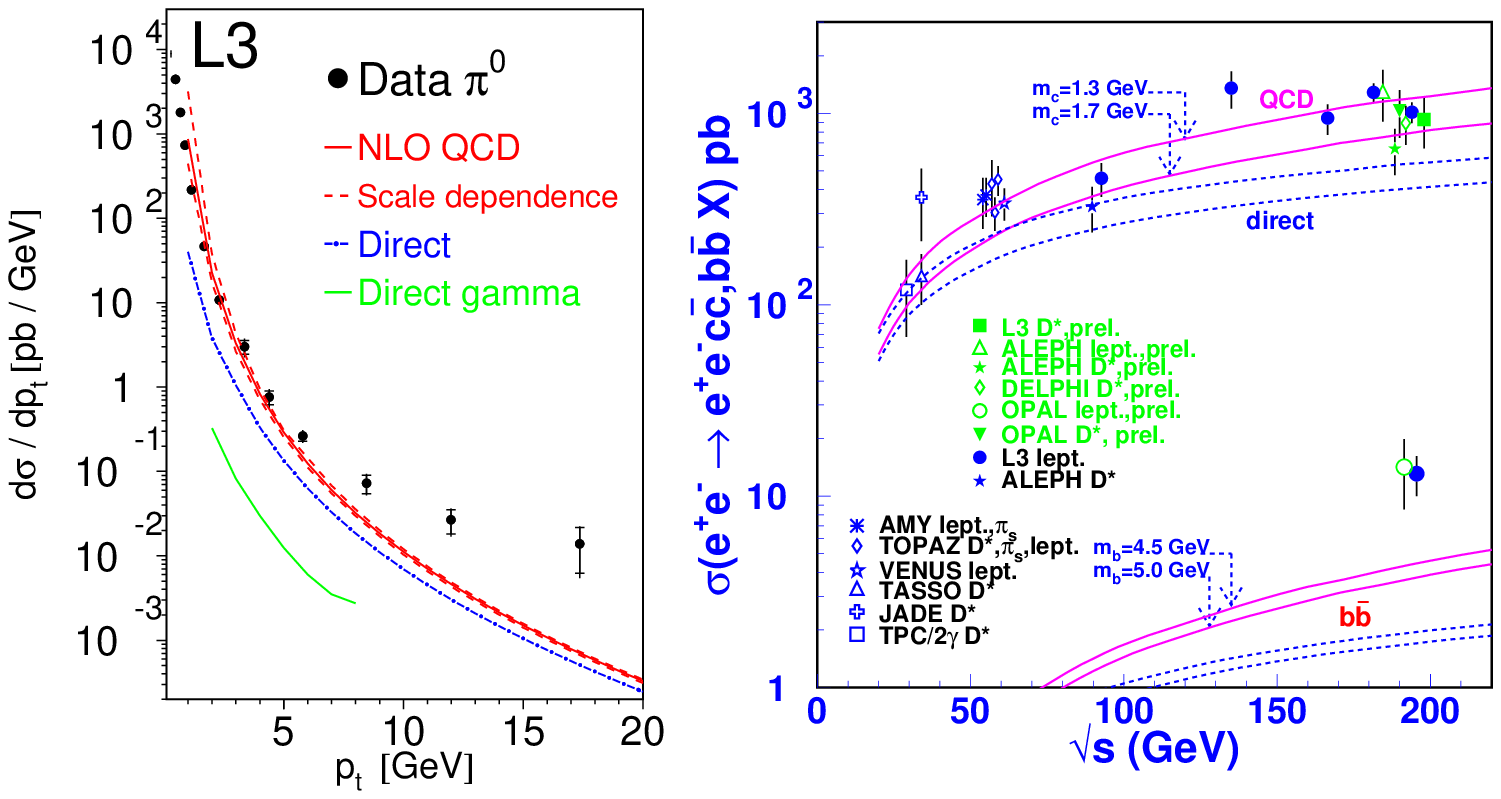}
\caption{Beauty cross-section as measured at HERA (left) and in  
$\gamma\gamma$ at LEP (right).} 
\label{fig:herabeauty}
\end{figure}

After the
first indication of
the excess
measured at the Tevatron,
also the HERA experiments found a similar 
excess both in photoproduction
and DIS events  (fig.~\ref{fig:herabeauty} left). Recently LEP experiments
reported again
a discrepancy between data and theory
for beauty production in 
$\gamma\gamma$ interactions  (fig.~\ref{fig:herabeauty} right). 
This situation is even more embarrassing if compared to what happens
in the case of central charm production, 
where QCD seems to work considerably better,
even if, due to the smaller mass of the $c$ quark, 
one should expect the contrary to happen
\footnote{It should be noted that the $b{\bar b}$ production cross section recently measured by HERA-B is instead in agreement with QCD predictions,
although the errors are large \cite{Abt:2002rd}.}.
\vspace*{0.5cm}

\section{Outlook}
With the advent of improved experimental techniques and
high-luminosity high-energy colliders more challenging QCD tests will become
possible. Nonetheless the role itself of the theory is going to change since
QCD will be the major source of background for a variety of processes.
More precise calculations and refined theoretical tools
are needed to perform stringent tests
but also to have a reliable control on QCD effects
in new physics scenarios.

\section{Acknowledgments}
We would like to thank the Organizers of
the IFAE meeting
for giving us the opportunity to convene this exciting working group.
Their
help and warm hospitality were 
certainly one of the main
ingredients of the success of the Workshop.

%
\end{document}